\documentclass[twocolumn,showpacs,preprintnumbers,amsmath,amssymb]{revtex4}

\usepackage{graphicx}
\usepackage{dcolumn}
\usepackage{bm}

\begin{document}


\title{Calculating Luminosity Distance versus Redshift in FLRW Cosmology via \\Homotopy
Perturbation Method}

\author{V. K. Shchigolev}
\altaffiliation[E-mail:]{ vkshch@yahoo.com}

\affiliation{
Department of Theoretical Physics, Ulyanovsk State University, 42 L. Tolstoy Str., Ulyanovsk 432000, Russia
}%

\date{\today}

\begin{abstract}{\bf Abstract} --- We propose an efficient analytical method for
estimating the luminosity distance
in a homogenous Friedmann-Lema\^itre-Robertson-Walker (FLRW) model of the Universe. This method is based on the
homotopy perturbation method (HPM), which has high accuracy in many nonlinear problems, and can be easily implemented.
For analytical calculation of the luminosity distance,  we offer to proceed not from the computation of the integral,
which determines it, but from the solution of a certain differential equation with corresponding initial conditions.
Solving this equation by means of HPM, we obtain the approximate analytical expressions for the luminosity distance as
a function of redshift for two different types of homotopy.
Possible extension of this method to other cosmological
models is also discussed.
\end{abstract}

\maketitle

\section{Introduction}
\label{intro}
Recent cosmological observations \cite{Riess}-\cite{Suzuki} clearly indicate that the present universe is a spatially
flat and expands with acceleration. The SNIa Union2 database gives us the reliable observational resources for testing
various cosmological models since the Supernovae of type Ia are one of the best cosmological distance indicators.
For this reason, the analytical calculation of the luminosity distance $d_L$ versus cosmological redshift $z$ becomes a very important
issue in theoretical cosmology.

As it is mentioned in \cite{Liu},
in order to undertake the comparison of any cosmological models with the type Ia Supernovae data,   an analytical approximation of the luminosity distance as a function of the redshift $z$ is required. The reason is that the corresponding formula for
the luminosity distance is usually expressed via an integral over the redshift, and the integration cannot be prepared explicitly.

In cosmology, it is quite common to encounter
physical quantities expanded as a Taylor series in the redshift (see, e.g. \cite{Cat}).
The most well-known example of this phenomenon is the Hubble
relation between luminosity distance and redshift. However, we now have
supernova data at least back to redshift $z \approx 2.3$ data available.
This opens up the theoretical question as to whether or
not such a series expansion actually converges for large redshift.
Therefore, there is a need to find other algorithms for computing the luminosity
distance as a function of redshift.

A simple algebraic approximation to
the luminosity distance $d_L$ and the proper angular diameter
distance  in a flat universe with pressureless
matter and a cosmological constant is presented in \cite{Pen}.

In \cite{Meszaros}, it was shown that the integral on the right hand side of general formula for the luminosity
distance can be partly calculated analytically using the elliptic
integral of the first kind even in the case, when all the three
omega factors are non-zeros. This calculation can be useful
for the certain restriction on the model parameter.

The so-called Pad$\acute{e}$ approximant was used in order to obtain the analytical approximation of the
luminosity distance for the flat XCDM model in \cite{Wei}. In order to simplify the repeated computation
of difficult transcendental functions or numerical
integrals, there were presented a fitting formula with some restricted properties.

At the same time, Dr. Ji-Huan He \cite{He} proposed an analytical method for solving differential
and integral equations, HPM, which is a combination of standard homotopy and the perturbation.
The HPM has a significant advantage in that it provides an analytical approximate solution to a wide range
of nonlinear problems in applied sciences. The application of the HPM  \cite{He2} - \cite{Cveticanin} includes
the nonlinear differential equations, nonlinear integral equations, fractional differential equations, and many others.
It has been shown that generally one or two iterations in this method can lead to highly accurate solutions.
The HPM yields a very rapid convergence of the solution series in most cases considered so far in literature.
Recently  there have been studies in which this method is used for analytical calculations in the field of cosmology
and astrophysics (see, e.g. \cite{Zare}-\cite{Rahaman}).

For analytical calculation of the luminosity distance, we offer to proceed  from the solution of differential equation
with certain initial conditions. Solving this equation in a spatially flat FLRW universe
by means of HPM, we are able to obtain the
approximate analytical expressions for the luminosity distance in terms of redshift for different types of homotopy.
We show that by using the homotopy perturbation method,
the explicit dependency $d_L(z)$ in arbitrary accuracy  can be easily obtained by
implementing a simple procedure for the governing equation.
Finally, we discus some possible extensions of HPM to other cosmological
models.

\section{Luminosity Distance Equation}
\label{sec:1}
We consider the FLRW metric \cite{Weinberg},
$$
ds^2=-dt^2+a^2(t)\Big[\frac{dr^2}{1-kr^2}+r^2(d\theta^2+\sin \theta d\phi^2)\Big],
$$
where $a(t)$ is a scale factor, and  $k=1, 0, -1$ for a closed, spatially-flat, open
universe respectively. Then, from the Einstein equations, one can obtain the Friedmann
equation in the following form
\begin{equation}\label{1}
H^2=H_0^2\big\{\Omega_r a^{-4}+\Omega_m a^{-3} + \Omega_k a^{-2}+\Omega_{\Lambda}\big\},
\end{equation}
where $H=\dot a/a$ is the Hubble parameter, $H_0$ denotes its present value, and we use dimensionless densities
$\Omega_M=(8\pi G/3H^2)\rho_M$, $\Omega_k=-k/H_0^2a_0^2$ and $\Omega_{\Lambda}=\Lambda/3H_0^2$.
Here, $\Omega_{\Lambda}$ is the
contribution from the vacuum, $\Omega_k$ is the contribution associated with curvature, and $\Omega_M$ is the contribution from all
other kinds of matter and fields.

As well known, the most fundamental distance scale in the universe is the luminosity
distance, defined by $d_L = \sqrt{L/(4\pi f )}$, where $f$ is the observed
flux of an astronomical object and $L$ is its luminosity. Recent astronomical
observations indicate that the present density parameter of
the universe satisfies $\Omega_M+\Omega_{\Lambda} = 1$ with $\Omega_{\Lambda}\sim 0.72$.
The distance calculations in such a vacuum-dominated
universe involve repeated numerical calculations and elliptic functions.

In order to simplify the numerical calculations,   a simple algebraic approximation to
the luminosity distance has been developed  to calculate the distances
in a vacuum-dominated flat universe \cite{Pen}, \cite{Wickramasinghe}.
In some cases, the general formula for the luminosity
distance can be partly calculated analytically using the elliptic
integral of the first kind. Nevertheless, the problem of analytical calculating the luminosity distance
remains interesting because of great amount of different models in which the Hubble parameter takes
more complicated dependence on $z$ than in equation (\ref{1}). We propose a novel approach to this problem
based on the homotopy perturbation method. For this purpose, we will first have to obtain the differential
equation which the luminosity distance should satisfy to, and define the appropriate initial conditions for this equation.

To verify any  cosmological model by the observational data, one should follow the maximum-likelihood approach
under which one minimizes $\chi^2$ and hence
measures the deviations of the theoretical predictions from the observations.
Since SN Ia behave as excellent standard candles, they can be used to directly measure the
expansion rate of the Universe upto high redshift, comparing with the present rate. The SN Ia data
gives us the distance modulus $\mu$ to each supernova.

In a flat universe, the theoretical distance modulus is given by
$$
\mu(z)=5 \log_{10}(d_L/Mpc)+25,
$$
where $d_L$ is the luminosity distance, and $p_s$ denotes model parameters.  For theoretical calculations,
the luminosity distance $d_L$ of SNe Ia is defined as follows
\begin{equation}\label{2}
d_L = c(1+z)\int\limits_o^z \frac{d z'}{H_0 E(z')},
\end{equation}
where $E(z)=H(z)/H_0$, and $H(z)$ is the Hubble parameter (\ref{1}), that is represented as a function of redshift.

For example, the luminosity distance
$d_L$ in a spatially flat $\Lambda$CDM universe is
given by \cite{Weinberg}
\begin{equation}
d_L(z)\! = \!\frac{c(1 + z)}{H_0}\!\!\!
\int\limits^{z}_{0}\!\!\!\frac{dz'}{\sqrt{\phantom{\dot A}\!\!\!\!\!\Omega_r(1\! +\! z' )^4+\!\Omega_m(1\! + \!z' )^3 + \!\Omega_{\Lambda}}},\label{3}
\end{equation}
where $\Omega_m$, $\Omega_r$ and $\Omega_{\Lambda}$ are the energy densities corresponding to the
matter, radiation and cosmological constant, respectively: $\Omega_m+\Omega_r + \Omega_{\Lambda}= 1$.

For the convenience of subsequent calculations, we introduce the following notation for the dimensionless
Hubble parameter squared
\begin{equation}\label{4}
E^2(z)=W(z),\,\,\, W(z)_{|\,z=0}=1.
\end{equation}
In the same example of the spatially flat $\Lambda$CDM universe,
\begin{equation}\label{5}
W(z)=\Omega_r(1 + z)^4+\Omega_m(1 + z)^3 + \Omega_{\Lambda}.
\end{equation}
Due to (\ref{4}), we can rewrite formula (\ref{2}) as follows
\begin{equation}\label{6}
d_L(z) = \frac{c(1 + z)}{H_0} \int\limits^{z}_{0} \frac{dz'}{\sqrt{\phantom{\dot A}W(z')}}.
\end{equation}
Expressing from the last equation  $H_0 d_L/c(1+z)$ as the integral in r.h.s., and differentiating result, we have
\begin{equation}\label{7}
\frac{d}{dz}\left[\frac{H_0 d_L}{c(1+z)}\right]=\frac{1}{\sqrt{\phantom{\dot A}W(z)}}.
\end{equation}
Then, the second derivative is equal to
$$
\frac{d^2}{dz^2}\left[\frac{H_0 d_L}{c(1+z)}\right]=-\frac{1}{2}W^{-3/2}(z)\frac{d W(z)}{d z}.
$$
Substituting $1/\sqrt{W(z)}$ from (\ref{7}) into the last equation, we obtain
\begin{equation}\label{8}
\frac{d^2}{dz^2}\left[\frac{H_0 d_L}{c(1+z)}\right]=-\frac{1}{2}\frac{d W(z)}{d z}\left(\frac{d}{dz}\left[\frac{H_0 d_L}{c(1+z)}\right]\right)^3.
\end{equation}
From equations (\ref{6}) and (\ref{7}), it is obvious that
\begin{equation}\label{9}
\left[\frac{H_0 d_L}{c(1+z)}\right]_{\Big| z=0}=0,\,\,\,\frac{d}{dz}\left[\frac{H_0 d_L}{c(1+z)}\right]_{\Big| z=0}=1.
\end{equation}

For the sake of simplicity, let us introduce
\begin{equation}\label{10}
x=1+z,\,\,\,\,u(x)=\frac{H_0 d_L}{cx}.
\end{equation}
With this and equations (\ref{8}), (\ref{9}), we get  the following Cauchy problem
\begin{equation}\label{11}
u''+\frac{1}{2}W'(x-1)u'^3=0\,;\,\,\,\,u_{\big| x=1}=0,\,u'_{\big| x=1}=1,
\end{equation}
where the prime stands for the derivative with respect to $x$, and $W(x-1)_{|\,x=1}=1$.

\section{Calculating Luminosity Distance via HPM }
\label{sec:2}

The main equation (\ref{11}) is a nonlinear differential equation of the second order. It can be solved exactly in quadratures,
but the result again leads to the formula (\ref{6}). Therefore, we will solve this equation analytically, but with
a certain approximation. Among all kinds of approximate methods we now use the HPM. In this method, it is not
required to introduce a small parameter, because it is naturally contained in the method itself.

Since the HPM has now become standard and for brevity, the reader is referred to \cite{He}-\cite{He7} for the basic ideas of HPM.
In this section, we shall apply the HPM
to solve equation (\ref{11}). Let us assume that the solution of this equation
can be represented by a series in $p$ as follows
\begin{equation}\label{12}
u=u_0+p\, u_1+p^2 u_2 + p^3 u_3 + ...\, .
\end{equation}
where $p \in [0, 1]$ is an imbedding parameter.
When we put $p \to 1$, then equation (\ref{3}) corresponds to (\ref{2}), and (\ref{5}) becomes the approximate solution
of (\ref{10}), that is
\begin{equation}\label{13}
u(x)= \lim_{p \to 1} u =u_0+u_1+ u_2 +  u_3 + ...\, .
\end{equation}
It is useful to note that the result of solving a nonlinear equation by this method and the convergence rate greatly
depend on the choice of the homotopy. Therefore, we consider two cases in what follows. In one case, we
construct the homotopy from the idea of simplicity. In the next case, we just follow  the procedure of the
general approach proposed in \cite{He}.

\subsection{Naive homotopy}

Applying this method to equation (\ref{11}) in this case, we build the following simplest homotopy:
\begin{equation}\label{14}
u''+p\frac{1}{2}\,W'(x-1)u'^3=0,\,\,\,\,p \in [0,1],
\end{equation}
and assume that this equation can be solved by means of the series in $p$ as (\ref{12}).

Substituting  (\ref{12}) into equation (\ref{14}), and equating coefficients of like powers of $p$,
one obtains the following equations:
\begin{eqnarray}
p^0&:&u\,''_0=0,   \nonumber\\
p^1&:&u\,''_1+\frac{1}{2}W'(x-1)u_0'^3=0,\nonumber\\
p^2&:&u\,''_2+\frac{3}{2}W'(x-1)u_0'^2u'_1=0, \label{15}\\
p^3&:&u\,''_3+\frac{3}{2}W'(x-1)(u'_0 u_1'^2+u_0'^2 u'_2)=0, \nonumber\\
& &.\,\,.\,\,.\,\,.\,\,.\,\,.\,\,.\,\,.\,\,.\,\,.\,\,.\,\,. \nonumber
\end{eqnarray}
According to (\ref{11}), the initial conditions for $u_i (x)$  can be chosen as follows
\begin{eqnarray}
u_{0\big| x=1}=0,\,\,\,\,u'_{0\big| x=1}=1;\nonumber\\
u_{j\big| x=1}=0,\,\,\,\,u'_{j\big| x=1}=0;\label{16}
\end{eqnarray}
where $j \geq 1$.

It is noteworthy that we obtain the set of linear equations (\ref{15}).
Their solutions with the initial conditions (\ref{16}) can be readily found as
\begin{eqnarray}
\nonumber  u_0(x) &=& x-1, \\
\nonumber  u_1(x) &=& -\frac{1}{2}\int\limits^{x}_{1}[W(x'-1)-1]dx', \\
\label{17}  u_2(x) &=& \frac{3}{8}\int\limits^{x}_{1}[W(x'-1)-1]^2dx', \\
\nonumber  u_3(x) &=& -\frac{5}{16}\int\limits^{x}_{1}[W(x'-1)-1]^3dx'.
\end{eqnarray}

Substituting  all solutions (\ref{17}) into equation (\ref{13}), we obtain
\begin{eqnarray}
u(x)=\int\limits^{x}_{1}\Big\{1\!\!\!&-&\!\!\!\frac{1}{2}[W(x'-1)-1]+\frac{3}{8}[W(x'-1)-1]^2 \nonumber\\
&-&\!\!\!\frac{5}{16}[W(x'-1)-1]^3+...\Big\}dx'.\label{18}
\end{eqnarray}
Then taking into account (\ref{4}) and (\ref{10}), we can express the luminosity distance in the following
form
\begin{eqnarray}
d_L(z)=\frac{c(1 + z)}{H_0}\int\limits^{z}_{0}\left\{1-\frac{1}{2}[W(z')-1]\right. \nonumber\\
\left.+\frac{3}{8}[W(z')-1]^2-\frac{5}{16}[W(z')-1]^3+...\right\}dz',\label{19}
\end{eqnarray}

The convergence of this solution is rather obvious, because formula (\ref{18})
could be obtained merely from the decomposition
$$
\frac{1}{\sqrt{W(z')}}=\frac{1}{\sqrt{1+(W(z')-1)}}=
$$
$$
=1-\frac{1}{2}[W(z')-1]+\frac{3}{8}[W(z')-1]^2
-\frac{5}{16}[W(z')-1]^3+...
$$
in equation (\ref{6}).

Let us consider an example. Substituting $E^2(z)$ for the spatially flat $\Lambda$CDM model from
(\ref{5}) into (\ref{19}), we obtain
\begin{eqnarray}
d_L=\frac{cz}{H_0}\left\{1+\frac{1}{2}\Big[2-2\Omega_r-\frac{3}{2}\Omega_m\Big]z ~~~~~~~~\right.\nonumber \\
\left.-\frac{1}{6}\Big[12\Omega_r+\frac{15}{2}\Omega_m -\frac{3(3\Omega_m+4\Omega_r)}{4}\Big]z^2+...\right\}, \label{20}
\end{eqnarray}
where we have used $\Omega_r+\Omega_m+\Omega_{\Lambda}=1$.
At the same time, the well known expansion of the luminosity distance $d_L$ in a Taylor series
in redshift $z$  yields (see, e.g. \cite{Chiba}-\cite{Visser2}):
\begin{equation} \label{21}
d_L(z) = \!\!\frac{cz}{H_0}\Big[1 +
\frac{1}{2}
(1 - q_0)z -\frac{1}{6}(1-q_0-3q_0^2+j_0)z^2 +O(z^3)
\Big],
\end{equation}
where the present magnitudes of the deceleration parameter $q=-a\ddot a/\dot a^2$ and the jerk parameter
$j=a^2\dddot a/\dot a^3$ are denoted as $q_0$ and $j_0$, respectively. Comparing equations (\ref{20}) and (\ref{21}),
one can get both $q_0$ and $j_0$. For example,
\begin{equation} \label{22}
q_0=-1+2\Omega_r+\frac{3}{2}\Omega_m=1-2\Omega_{\Lambda}-\frac{1}{2}\Omega_m.
\end{equation}

Not pretending here on the observational constraints of this model but only with the illustrative purpose,
let us put $\Omega_{\Lambda}=0.72$ and $\Omega_m \approx0.28$ into (\ref{22}). Then we have $q_0\approx-0.58$, that indicates
the accelerated expansion. It is quite clear that, owing to its simplicity, formula (\ref{19}) may be useful in testing a variety
of cosmological models via the observational data .

\subsection{Enhanced homotopy}

In this case, we build the homotopy according to the general procedure of the method, namely
\begin{equation} \label{23}
u''+\frac{W'(0)}{2}+p\Big[\frac{1}{2}W'(x-1)u'^3-\frac{W'(0)}{2}\Big]=0,
\end{equation}
where $p \in [0,1]$, and
\begin{equation} \label{24}
W'(0)=\frac{dW(x-1)}{dx}_{\Big|\,x=1}=\frac{dW(z)}{dz}_{\Big|\,z=0}
\end{equation}
is a constant followed from equation (\ref{4}). The term $W'(0)$ is introduced into equation (\ref{23}) due to equation (\ref{11})
taken at $x=1$.
\begin{figure}[thbp]
\centering
\includegraphics[width=0.45\textwidth,height=0.4\textwidth]{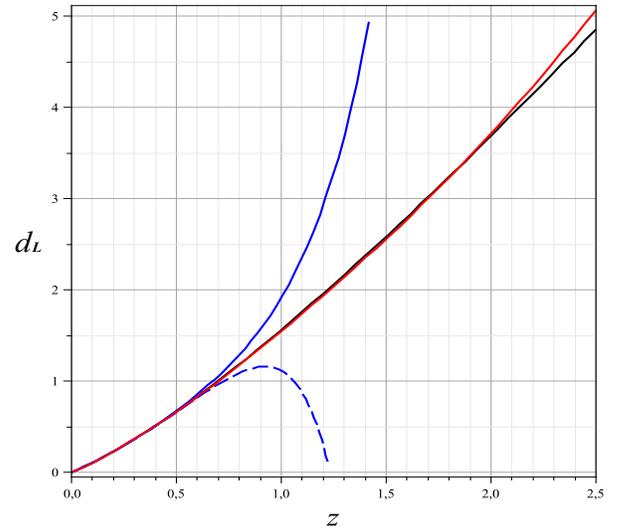}
\caption{Comparison of the approximate solutions, given by  Eq. (19)
(blue line to the second iteration, and blue dashed  line to the third one)
and Eq. (29) (red line) , with the numerical solution to Eq. (6) (black line).}
\label{Figure_1}
\end{figure}

In the case of homotopy  (\ref{23}) , substituting  (\ref{12}) into equation (\ref{23}) and equating coefficients of like powers of $p$,
we obtain the following set of equations:
\begin{eqnarray}
p^0&:&u\,''_0+\frac{W'(0)}{2}=0, \label{25}  \\
p^1&:&u\,''_1+\frac{1}{2}\Big[W'(x-1)u'^3_0-W'(0)\Big]=0,\label{26}\\
& &.\,\,.\,\,.\,\,.\,\,.\,\,.\,\,.\,\,.\,\,.\,\,.\,\,.\,\,. \nonumber
\end{eqnarray}
where we have stopped at the first iteration for the sake of simplicity.
All subsequent approximations can also be obtained easily.

Solving equation (\ref{25}) with the initial conditions (\ref{16}), one can get
\begin{equation} \label{27}
u_0=(x-1)- \frac{W'(0)}{4}(x-1)^2.
\end{equation}
Substituting this function into equation (\ref{26}), and taking into account (\ref{16}), we
obtain
\begin{eqnarray} \label{28}
u_1=\frac{1}{2}(x-1)+ \frac{W'(0)}{4}(x-1)^2 \nonumber \\
-\frac{1}{2}\int\limits^{x}_{1}W(x'-1)\Big[1-\frac{W'(0)}{2}(x'-1)\Big]^2 \nonumber \\
\times\Big\{1-\frac{W'(0)}{2}[4(x'-1)-3(x-1)]\Big\} dx',
\end{eqnarray}
where we have used the Cauchy formula for the m-fold integral.

With the accuracy accepted in this case, we obtain the approximate solution as
\begin{eqnarray} \label{29}
d_L=\frac{c(1+z)}{H_0}\Big\{\frac{3}{2}z-\frac{1}{2}\times ~~~~~~~~~~~~~~~~~~\nonumber \\
\int\limits^{z}_{0}\!\!W(z')\Big(1\!-\!\frac{W'(0)}{2}z'\Big)^2\!\Big[1\!-\!\frac{W'(0)}{2}(4z'\!-\!3z)\Big] dz'\Big\},
\end{eqnarray}
by adding (\ref{27}) and (\ref{28}), and taking into account definitions (\ref{10}).

Substituting again the expression for $W(z)$ from equation (\ref{5}) into the formula (\ref{29}), one can verify that, within
the same accuracy, equation (\ref{29}) yields the same result (\ref{20}). However, it is noteworthy that here this result
has been obtained just by the single iteration. Furthermore, accounting terms of the expansion in $z$ of a higher order in
equations (\ref{19}) and (\ref{29}) shows even more fundamental difference between them.

Using the Maple package, we get
the graphs of $d_L(z)$ (in units of $c/H_0$) for the numerical solution to the integral in the r.h.s. of equation (\ref{6}), and the approximate
solutions obtained by the naive homotopy, Eq. (\ref{19}), and by the enhanced homotopy, Eq. (\ref{29}), shown in Figure 1.
In all these cases, we have used $\Omega_m=0.28, \Omega_{\Lambda}=0.72,\Omega_r=0$, as an illustrative example.
Table 1 shows the percentage of relative errors of the approximate solutions compared to the numerical one for the same example.

\begin{quote}
\textbf{Table 1:} Percentage of relative errors of the approximate solutions to $d_L$ in cases of Eqs. (\ref{19}) and (\ref{29}).
\end{quote}
\begin{center}
\begin{tabular}
[c]{|l||l|l|}\hline
\vspace{-3mm} \, & \, & \,\\
$z$ & Errors \,\%\,\,by\,Eq.(19) & Errors\, \%\,\,by\,Eq.(29)\\\hline \hline
\vspace{-3mm} \, & \, & \,\\
0.1 & 0.00362 & 0.00164\\\hline
\vspace{-3mm} \, & \, & \,\\
0.3 & 0.04995 & 0.04243\\\hline
\vspace{-3mm} \, & \, & \,\\
0.5 & 0.62629 & 0.17769\\\hline
\vspace{-3mm} \, & \, & \,\\
0.7 & 3.74942 & 0.44171\\\hline
\vspace{-3mm} \, & \, & \,\\
1.0 & 28.45987 & 0.86172\\\hline
\vspace{-3mm} \, & \, & \,\\
1.2 & -- & 1.08054\\\hline
\vspace{-3mm} \, & \, & \,\\
2.0 & -- & 0.66687\\\hline
\end{tabular}
\end{center}

\section{Discussion}

The results of the preceding section clearly demonstrate the advantage of the formula (\ref{29}) in comparison with approximation (\ref{19}).
Obviously, a better approximation to the exact magnitude of the luminosity distance could be reached by the second iteration in the case of
enhanced homotopy.
For this end, as one can see from the equation (\ref{23}),  we  have to solve the following equation
$$
u\,''_2+\frac{3}{2}W'(x-1)u_0'^2u'_1=0,
$$
given that $u_2(1)=u'_2(1)=0$, and
$$
u'_0 = 1-\frac{W'(0)}{2}(x-1),
$$
$$
u'_1=\frac{1}{2}\Big\{1+W'(0)(x-1)-W(x-1)\Big[1-\frac{W'(0)}{2}(x-1)\Big]^3\Big\}.
$$
It is not a difficult problem to solve this linear equation in quadratures, if required for the greater accuracy.
Nevertheless, from Table 1  it can be concluded  that  the relative error of the simple approximation by the formula
(\ref{29})) is mostly less then 1\% for the redshift within the interval from 0 to  2, or even greater.

It would be noted that our method can be extended to cover the non-flat  $\Lambda$CDM models with a
curvature term $\Omega_k$ in (\ref{1}), because in that case the equivalent of equation (\ref{3})
assumes the form
$$
d_L=\frac{c(1+z)}{H_0\sqrt{|\Omega_k|}}\sin_k\left[\sqrt{|\Omega_k|}\int\limits^{z}_{0} \frac{d z'}{E(z'; \Omega_i)}\right],
$$
where
$$E^2(z; \Omega_i)=
\Omega_r(1 + z)^4+\Omega_m(1 + z)^3 + \Omega_k(1 + z)^2 + \Omega_{\Lambda},
$$
with $\Omega_r+\Omega_m + \Omega_k + \Omega_{\Lambda}=1$. The $\sin_k(x)$ function
is defined by
$$\sin_k(x) = \left\{
\begin{array}
[c]{rcl}
\sin(x), ~~~\Omega_k < 0,  &  & \\
\sinh(x)~~~\Omega_k > 0. &  &
\end{array}
\right.
$$
Let us denote the inverse function to $\sin_k(x)$ as  $\arcsin_k(x)$. Assuming the following notation
$$
u(x)=\frac{1}{\sqrt{|\Omega_k|}}\arcsin_k\left(\frac{\sqrt{|\Omega_k|}H_0 d_L}{cx}\right)
$$
instead of (\ref{10}), one can obtain the main equation
for the luminosity distance, coinciding with equation (\ref{11}).
After that, a little change, for example, of formula (29) for $d_L$ becomes obvious:
The multiplier $\displaystyle \frac{c(1+z)}{H_0}$\Big\{...\Big\} should be replaced by $\displaystyle \frac{c(1+z)}
{H_0\sqrt{|\Omega_k|}}\sin_k\Big\{\sqrt{|\Omega_k|}...\Big\}$.

However, the most important conclusion that follows from an analysis of the approach developed in this work is as follows.
The approach and, for example,  formula (\ref{29}) can be applied not only to the $\Lambda$CDM models, but also to any other FLRW model.
This approximation is  especially useful in those cases where the Hubble parameter squared is not a polynomial in $(1+z)$, and
the computation of the integral in (\ref{2}) becomes problematic at all. The proposed approach allows to get an analytic
expression for the function $d_L(z)$ with a high accuracy  without any problem in most cases, at least by using the Maple package
for calculation of the integral in (\ref{29}).

\section{Conclusions}
\quad
Thus, in this paper,  we have considered a simple analytical computation of the luminosity distance in General Relativity by means of
the  Homotopy Perturbation Method. For this purpose, we first have transformed the problem of calculating the integral
in the expression for the luminosity distance (\ref{2}) to the problem of solving the Cauchy problem for the corresponding nonlinear
differential equation (\ref{11}).
Thereafter, the resulting equation has been solved with the help of the approximate analytic method, namely HPM.
Two different choices of homotopy, Eqs. (\ref{14}) and (\ref{23}), were considered, and all  solutions were obtained in quadratures.
Thus, in this paper, we have obtained the new analytical approximations for the luminosity distance.

The comparison of our solutions with the corresponding numerical solution for a flat $\Lambda$CDM model (see Fig. 1) clearly showed a high
accuracy of the HPM approximation, at least for the redshift less then 2.
The obvious advantage of the formula obtained is the fact that it does not initially involves a Taylor series expansion
over the redshift, that is over the integer powers of redshift.

Unfortunately, this method is sensitive to the choice of  homotopy. The method does not give us strong recommendation
how to make the best choice among the unbounded number of different possibilities. In the first case, we have intentionally
used the simplest homotopy in order to just show the main steps in obtaining an approximate solution by this method.
At the same time, we have got almost obvious result that allows us to demonstrate the convergence of the method.

There exist alternative approaches to the construction of homotopy. The second example of homotopy shows that even a few number
of iteration steps leads to a high accuracy.
So, it can be concluded that the HPM is a powerful and efficient technique to solve the problem of the luminosity
distance computation in theoretical cosmology.


\begin{thebibliography}{99}
\bibitem{Riess} A. G. Riess, et al.,
Astron. J.  {\bf116}, 1009 (1998).

\bibitem{Perlmutter} S. Perlmutter, et al.,  Astrophys. J.  {\bf517},  565 (1999).

\bibitem{Suzuki} N. Suzuki,  D. Rubin, C. Lidman,  et al.,  Astrophys. J. {\bf746}, 85 (2012).

\bibitem{Liu} De-Zi Liu, Cong Ma, Tong-Jie Zhang, and Zhiliang Yang, Mon. Not. R. Astron. Soc. {\bf412}, 2685 (2011).

\bibitem{Cat} C$\acute{e}$line Catto$\ddot e$n and Matt Visser,  Class. Quantum Grav. {\bf24}, 5985 (2007).

\bibitem{Pen} Ue-Li Pen, Astrophys. J. Suppl. S. {\bf120}, 49È50 (1999).

\bibitem{Meszaros} A. Meszaros, J. Ripa,  Astron. Astrophys. {\bf573},  A54 (2015).

\bibitem{Wei} Hao Wei, Xiao-Peng Yan, Ya-Nan Zhou,  JCAP {\bf1401}, 045 (2014).

\bibitem{He} J.-H. He, Comput. Meth. Appl. Mech. Eng. {\bf178}, 257 (1999).

\bibitem{He2} J.-H. He,  Int. J. Nonlinear Mech. {\bf35}(1),  37 (2000).

\bibitem{He3} J.-H. He, Appl. Math.
Comput.  {\bf135}, 73 (2003).

\bibitem{He6} J.-H. He, Indian J.
Phys.  {\bf88}(2), 193  (2014).

\bibitem{He7} J.-H. He, Abstract and Applied Analysis
{\bf2012}, Article ID 916793, 130 pages;
DOI:10.1155/2012/916793

\bibitem{Cveticanin} L. Cveticanin,
Chaos Soliton Fract. {\bf30}(5), 1221 (2006).

\bibitem{Zare} M. Zare, O. Jalili and M. Delshadmanesh,
Indian. J. Phys.   {\bf86}(10), 855 (2012).

\bibitem{Shchigolev1} V.  Shchigolev, Universal Journal of Applied Mathematics {\bf2}(2), 99 (2014).

\bibitem{Rahaman}F. Rahaman, S. Ray, A. Aziz, S. R. Chowdhury, D. Deb, arXiv:1504.05838.


\bibitem{Weinberg}  S. Weinberg,  {\it Gravitation and Cosmology: Principles and Applications of The
General Theory of Relativity} (John Wiley. Press, New York, 1972).

\bibitem{Wickramasinghe} T. Wickramasinghe and T. N. Ukwatta, Mon. Not. R. Astron. Soc. {\bf406} , 548   (2010).

\bibitem{Chiba} Takeshi Chiba, Takashi Nakamura,  Prog. Theor. Phys. {\bf100},  1077 (1998).


\bibitem{Visser1} M. Visser,  Class. Quantum Grav. {\bf21}, 2603 (2004).

\bibitem{Visser2} M. Visser,  Gen. Rel. Grav. {\bf37}, 1541 (2005).

\end{thebibliography}

\end{document}